\documentclass[prb,aps,eqsecnum]{revtex4}
\usepackage{graphicx}
\newcommand{\e}{{\rm e}}
\newcommand{\ep}{\varepsilon}

\newcommand{\be}{\begin{equation}}
\newcommand{\ee}{\end{equation}}
\newcommand{\ba}{\begin{eqnarray}}
\newcommand{\ea}{\end{eqnarray}}
\newcommand{\bea}{\begin{eqnarray}}
\newcommand{\eea}{\end{eqnarray}}

\newcommand{\nn}{\nonumber}
\newcommand{\la}{\label} 
\newcommand{\w}{\omega} 
\newcommand{\hT}{T}
\newcommand{\hV}{V}

\newcommand{\htH}{H_A}

\def\t1{e_{_T}}
\def\v1{e_{_V}}
\def\ct{e_{_{TTV}}}
\def\cv{e_{_{VTV}}}
\def\tt{e_{_{TTTTV}}}
\def\tv{e_{_{VTTTV}}}
\def\vt{e_{_{TTVTV}}}
\def\vv{e_{_{VTVTV}}}

\begin{document}
\title{Gradient Symplectic Algorithms for Solving the
Radial Schr\"odinger Equation}

\author{Siu A. Chin and Petr Anisimov}
\affiliation{Department of Physics, Texas A\&M University,
College Station, TX 77843, USA}

\begin{abstract}

The radial Schr\"odinger equation for a spherically symmetric
potential can be regarded as a one dimensional classical 
harmonic oscillator with a time-dependent spring constant.
For solving classical dynamics problems, symplectic integrators
are well known for their excellent conservation properties.
The class of {\it gradient} symplectic algorithms is 
particularly suited for solving harmonic oscillator dynamics.
By use of Suzuki's rule for decomposing time-ordered operators, 
these algorithms can be easily applied to the  
Schr\"odinger equation. 
We demonstrate the power of this class of gradient algorithms 
by solving the spectrum of highly singular radial potentials 
using Killingbeck's method of backward Newton-Ralphson iterations.

\end{abstract}
\maketitle

\section {Introduction}
Because of its physical importance,
an immense literature exists for solving the radial Schr\"odinger equation,
\be
\frac{d^2 u(r)}{dr^2}=f(r,E)\,u(r),
\la{rad}
\ee
where
\be
f(r,E)=2V(r)-2E +\frac{\ell(\ell+1)}{r^2}.
\ee
This is usually solved by finite difference methods, such as the 
well known fourth-order Numerov\cite{hartree} algorithm, or further improved
schemes\cite{rchen}. Recent devlopments\cite{raptis,simos,vyver} have
resulted in many exponentially fitted algorithms which seek to
integrate (\ref{rad}) exactly when $f$ is a constant. As we
will see, because $f$ can vary rapidly with $V(r)$, specially in the
case of singular potentials, these algorithms do not in general
perform better than non-fitted algorithms.  
 
If we relabel the variables $r\rightarrow t$ and $u\rightarrow q$, (\ref{rad})
is just a 1D harmonic oscillator with a time-dependent spring constant 
$k(t,E)=-f(t,E)$,
\be
\frac{d^2 q(t)}{dt^2}=-k(t,E)\,q(t).
\la{rad2}
\ee
The novel aspect of the problem is that $k(t,E)$ can change sign with time
and become repulsive beyond the turning-point.	The dynamics of (\ref{rad2}) corresponds to 
a Hamiltonian with an explicit time-dependent potential,
\be
H=\frac12 p^2+\frac12 k(t,E)\,q^2.
\la{ham}
\ee
Thus any algorithm that can solve the classical 
time-dependent force problem
can be used to solve the radial Schr\"odinger equation. For example, one can
use Runge-Kutta type algorithms\cite{battin}. However, for solving 
classical dynamics, symplectic integrators\cite{dragt,yoshi,mcl02,hairer} are  
algorithms of choice because they conserve all Poincar\'e invariants and are 
deeply rooted in the Poisson formulation of classical mechanics. For oscillatory 
problems, symplectic algorithms are known to conserve energy and reduce phase 
error much better than Runge-Kutta type algorithms\cite{gla91,chin97,chinkid,chinchen03,chinsante}. 
The difficulty here is that, in order to derive an algorithm for solving
time-dependent dynamics, one must solve the problem of 
time-ordered exponential. Liu {\it et al.}\cite{liu} have recognized 
the time-dependent Hamiltonain structure of the Schr\"odinger 
equation, but were able to solve the time-dependent exponential,
and devised a symplectic algorithm, only to second order. Kalogiratou, Monovasilis 
and Simos\cite{kalog} have proposed a third order symplectic algorithm by 
expanding out the exponential to third order. Such a brute force approach 
cannot be extended to higher orders. A more systematic way of dealing with 
the time-ordered exponential is via the Magnus expansion\cite{wensch,baye,gold}, but the 
Magnus expansion requires explicit time integration in addition to evaluating 
higher order commutators. A more elegant solution is Suzuki's\cite{suz93} 
reinterpretation of the time-order exponential as reviewed in 
Ref.\cite{chinchen02}. By adapting Suzuki's rule, any factorized symplectic 
algorithms can be used to solve problems with an explicit time-dependent 
potential\cite{chinchen02}, including the disguised
radial Schr\"odinger equation (\ref{rad2}).  

     In order to devise efficient algorithms for solving the
radial Schr\"odinger equation (\ref{rad}), one must take 
advantage of its harmonic oscillator character (\ref{rad2}).
Most algorithms, even factorized symplectic ones, 
are general purpose algorithms and are not specially
adapted for solving the time-dependent harmonic oscillator. 
However, the recent class of {\it gradient} symplectic
algorithms\cite{chin97,chinkid,chinchen02,chinsante,chinchen03,ome02,ome03}, 
while general, seem tailor-made for solving
harmonic type dynamics. This is because  
these algorithms require computing the 
force gradient in addition to the force.
While the force gradient is not difficult to compute, 
it is especially trivial in the case of the harmonic oscillator.
This class of gradient (or more specifically, {\it forward}) integrators, 
has been shown to be efficient in solving both 
classical\cite{chin97,chinkid,chinchen03,chinsante,ome02,ome03} and 
quantum\cite{chinchen02,baye,gold,chin2001,chinkro} dynamical problems.
In this work, we will show they are also ideally suited for solving the 
radial Schr\"odinger equation. 

In the next section, we briefly summarize the Lie-Poisson operator 
formulation of symplectic integrators and Suzuki's rule for factorizing 
time-ordered exponentials. In Section III, we describe
forward, gradient based symplectic algorithms. In Section IV we review 
Killingbeck's method\cite{kill85} of eigenvalue-function determination. 
In Section V, we compare results on the Coulomb and other
singular radial potentials.
In Section VI, we discuss the applicability
of sixth-order algorithms and draw some 
conclusions in Section VII. 

\section {Time-dependent symplectic algorithms}
The Poisson bracket for evolving any dynamical variable $W(q,p)$
can be regarded as an operator equation
\begin{equation}
{{d }\over{dt}}W(q,p)=\{W,H\}\equiv
          \Bigl(
                  {{\partial H}\over{\partial p}}
				  {{\partial  }\over{\partial q}}
                 -{{\partial H}\over{\partial q}}
                  {{\partial  }\over{\partial p}}
				                    \Bigr)W,
\label{peq}
\end{equation}
with formal solution
\begin{equation}
W(t+\ep)={\rm e}^{\ep(T+V)}W(t).
\label{peqform}
\end{equation}
For the standard Hamiltonian, 
\begin{equation}
H(p,q)=\frac12 p^2+V(q),
\label{haml}
\end{equation}
the operators $T$ and $V$ are first order 
differential operators
\begin{equation}
T={{\partial H}\over{\partial p}}
				  {{\partial  }\over{\partial q}}
=p{{\partial }\over{\partial q}},\qquad
V=-{{\partial U}\over{\partial q}}
                  {{\partial  }\over{\partial p}}
=F(q){{\partial }\over{\partial p}}.
\label{tandv}
\end{equation}
The Lie transforms\cite{dragt} ${\rm e}^{\ep T}$ and ${\rm e}^{\ep V}$, are then
{\it displacement} operators which displace $q$ and $p$ forward in time via  
\begin{equation}
q\rightarrow q+\ep p\qquad{\rm and}\qquad 
p\rightarrow p+\ep F.
\label{peqv}
\end{equation}
Each factorization of ${\rm e}^{\ep (T+V)}$ into products  
of ${\rm e}^{\ep T}$, ${\rm e}^{\ep V}$ (and exponentials
of commutators of $T$ and $V$) gives rise to a 
{\it symplectic} algorithm, which is a sequence of successive
displacements (\ref{peqv}) 
for evolving the system forward in time. This is the fundamental
Lie-Poisson theory of symplectic integrators which has been studied 
extensively in the literature\cite{dragt,yoshi,mcl02,hairer}.

For a time-dependent Hamiltonian
\begin{equation}
H(t)=\frac12 p^2+V(q,t),
\label{hamt}
\end{equation}
the solution is given by the time-order exponential
\begin{equation}
W(t+\ep)=T\exp\Bigl(\int_t^{t+\ep}[\,T+V(s)\,]ds\Bigr)W(t),
\label{expth}
\end{equation}
where $V(t)$ is now the explicitly time-dependent operator
\be
V(t)=-{{\partial U(q,t)}\over{\partial q}}
                  {{\partial  }\over{\partial p}}
=F(q,t){{\partial }\over{\partial p}}.
\ee
Suzuki proved\cite{suz93} that 
\begin{equation}
T\exp\Bigl(\int_t^{t+\ep}[\,T+V(s)\,]ds\Bigr)=\e^{\ep(T+V(t)+D)},
\label{tdecom}
\end{equation}
where $D$ is 
the {\it forward time derivative} operator
\begin{equation}
D={{\buildrel \leftarrow\over\partial}\over{\partial t}}
\label{ftsh}
\end{equation}
such that for any two time-dependent functions $F(t)$ and $G(t)$,
\begin{equation}
F(t){\rm e}^{\ep D}G(t)=F(t+\ep)G(t).
\label{fg}
\end{equation}
Thus symplectic algorithms for solving explicitly time-dependent problems
of the form (\ref{hamt}) can be obtained by factorizing the three-operator
exponential of (\ref{tdecom}). Since $D$ commutes with $T$, one can first 
group 
\be
\widetilde T=T+D
\la{ttil}
\ee 
and factorize $\widetilde T$ and $V(t)$ as in the
time-independent case. The difference between time-dependent
and time-independent algorithms resides solely in the use of
$\widetilde T$ in place of $T$. This makes it extremely easy
to analyze and devise time-dependent algorithms.
Once factorized in terms of 
${\rm e}^{\ep\widetilde T}={\rm e}^{\ep T}
{\rm e}^{\ep D}$, the operator ${\rm e}^{\ep D}$ 
then shifts the time at which all the time-dependent 
potential to its LEFT must be evaluated. This results in Suzuki's 
rule for solving time-dependent Hamiltonian (\ref{hamt}):
{\it the time-dependent potential must be evaluated at an 
incremental time from the start of the algorithm 
equal to the sum of time steps of all the $T$ operators to its 
right}. We will illustrate how this is applied in the next 
section. For more detailed discussions and examples, see Refs.\cite{chinchen02,chinchen03}.     

\section {Forward fourth-order algorithms}

     In order to solve the radial Schr\"odinger equation (\ref{rad})
efficiently, one must take advantage of its harmonic oscillator character 
(\ref{rad2}). This can be done easily for factorized algorithms because 
their error terms have a well-defined analytical structure.
Consider the second order factorization,
\be
{\rm e}^{{1\over 2}\ep T}{\rm e}^{\ep V}
{\rm e}^{{1\over 2}\ep T}
=\exp\ep\Bigl[(T+V)
+{1\over{24}}\ep^2([T,[V,T]]-2[V,[T,V]])+O(\ep^4)\,\Bigr].
\label{sym2}
\ee
This is just a general operator equality which follows from the
Baker-Campbell-Hausdorff (BCH) formula. In the present context,
this equlity tells us that the second-order 
factorization on the LHS deviates from the exact evolution operator 
$\exp^{\ep(T+V)}$ by error terms which are the double 
commutators on the RHS. However, for the ordinary harmonic oscillator 
Hamiltonian (\ref{ham}) with a {\it constant} spring 
constant $k=\w^2$, one can easily verify that
\be
[V,[T,V]]=-2\w^2V,
\la{vtv}
\ee
\be 
[T,[V,T]]=-2\w^2T.
\la{ttv}
\ee
Thus the error terms can be re-expressed in terms of the
original operator $T$ and $V$ and be 
moved back to the LHS to yield,
\be
{\rm e}^{\ep(\frac12+\frac1{24}\w^2\ep^2) T}
{\rm e}^{\ep(1-\frac16 \w^2\ep^2) V}
{\rm e}^{\ep(\frac12+\frac1{24}\w^2\ep^2) T}
=\e^{\ep(T+V+O(\ep^4))}.
\label{sym22}
\ee
This means that the LHS is now
a fourth-order algorithm for solving the harmonic oscillator.
Because of the fundamental identities (\ref{vtv}) and 
(\ref{ttv}), all higher order 
commutators for the harmonic oscillator can be 
subsummed back to $T$ and $V$ yielding the exact
factorization\cite{chinkro}
\be
\e^{\ep C_E T}
   \e^{\ep C_M V}\e^{\ep C_E T}=\e^{ \ep(T+V)},
\la{exfact}
\ee
where the ``edge" coefficeint $C_E$ 
and the ``middle" coefficient $C_M$ are given by
\be 
C_E=\frac{1-\cos(\w \ep)}{\w\ep\sin(\w\ep)}
\quad{\rm and}\quad
C_M=\frac{\sin(\w\ep)}{\w \ep}.
\la{ctr}
\ee
The above discussion only depends on the abstract
commutator relations (\ref{vtv}) and (\ref{ttv}), and is
independent of the specific form of the operator $T$ and $V$.
Thus by interchanging $T\leftrightarrow V$, we can also factorize 
exactly,
\be
\e^{ \ep(T+V)}=\e^{\ep C_E V}
   \e^{\ep C_M T}\e^{\ep C_E V}.
\la{exfact2}
\ee

To solve the time-dependent harmonic oscillator,  
one has to replace $T\rightarrow \widetilde T$ everywhere.	
It is easy to verify that for any two
time-dependent functions $W(t)$ and $V(t)$,
\be
[V(t),[D,W(t)]]=0.
\la{dcom}
\ee
Hence, the commuatator (\ref{vtv})
\be
[V,[\widetilde T,V]]=[V,[T,V]]=2f(t,E)V
\la{tdvtv}
\ee
remains proportional to $V$. However,
\be
[\widetilde T,[V,\widetilde T]]=2f(t,E)T
+2[T,\frac{\partial}{\partial t}V]
-\frac{\partial^2}{\partial t^2}V
\la{tdtvt}
\ee
bears no simple relationship to $\widetilde T$. This means that one can still
retain the commutator $[V,[\widetilde T,V]]=[V,[T,V]]$ and move it back to
the LHS, but in order to have a fourth-order algorithm, one must
eliminate the commutator $[\widetilde T,[V,\widetilde T]]$ by more elaborated
factorization schemes. This coincides precisely with the way {\it forward}
symplectic algorithms are derived\cite{suzfour, chin97, chinchen03}. For
example, the simplest fourth-order forward factorization 
scheme\cite{suzfour,chin97} 
4B for evolving the system forward for time $\ep$ is
\ba
{\cal T}_{B}^{(4)}(\ep)
&=&
  {\rm e}^{a\ep \widetilde T}
  {\rm e}^{ \ep {1\over 2} V^* } 
  {\rm e}^{b\ep \widetilde T}
  {\rm e}^{ \ep {1\over 2} V^* } 
  {\rm e}^{a\ep \widetilde T},
  \nn\\
&=&
  {\rm e}^{a\ep T}
  {\rm e}^{ \ep {1\over 2} V^*(t+a^\prime\ep) } 
  {\rm e}^{b\ep T}
  {\rm e}^{ \ep {1\over 2} V^*(t+a\ep) } 
  {\rm e}^{a\ep T},
\la{fourb}
\ea 
where 
$a={1\over 2}(1-{1\over{\sqrt 3}})$,
$b={1\over{\sqrt 3}}$, $a^\prime=a+b={1\over 2}(1+{1\over{\sqrt 3}})$,
and the effective potential operator is given by
\ba
V^*(t)
&&=V(t)+{1\over 24}(2-\sqrt 3)\ep^2[V(t),[\widetilde T,V(t)]]
\nonumber\\
&&=\Bigl[1+{1\over 12}(2-\sqrt 3)\ep^2 f(t,E)\Bigr]V(t).
\label{superv}
\ea
This results in the use of an effective time-dependent
force
\be
F^*(t,E)=[1+{1\over 12}(2-\sqrt 3)\ep^2 f(t,E)\Bigr]f(t,E)q,
\la{efff}
\ee
which is no more difficult to evaluate than the original.
Factorization scheme (\ref{fourb}) translates into the following
fourth-order algorithm
for solving the time-dependent harmonic oscillator:
\ba
{ q}_1&=&{ q}_0+a\ep\,{ p}_0\, \nn\\
{ p}_1&=&{ p}_0+\frac12 \ep F^*(t+a\ep) q_1\,\nn\\ 
{ q}_2&=&{ q}_1+b\ep\,{ p}_1\, \la{alg4b}\\
{ p}_2&=&{ p}_1+\frac12 \ep F^*(t+a^\prime\ep) q_2\,\nn\\ 
{ q}_3&=&{ q}_2+a\ep\,{ p}_2\, \nn
\ea
The last numbered $p$ and $q$ are the updated values.
In the present context, since $q$ is the wave function and
$p$ is only an ancillary variable, we will be interested only
in algorithms that begins and ends with $q$. These 
{\it position}-type algorithms make full use of 
force evaluations at intermediate time to update the final position.
As will be discuss in the next Section, this 
point is important for Killingbeck's method of iterating the
last position $q(0,E)$ to zero.

In general, the commutator
\begin{equation}
[V,[T,V]]=2F_j{ {\partial  F_i}\over{\partial q_j} }
                { {\partial  }\over{\partial p_i} }=
                \nabla_i(|{\bf F}|^2)
                { {\partial  }\over{\partial p_i} }
\label{vtvf}
\end{equation}
produces a force which is the gradient of the square of the
original force. For the 1D harmonic oscillator, 
this is simply $\partial (fq)^2/\partial q=2f^2q$.
By incorporating the force gradient, algorithm 4B (\ref{alg4b}) 
is fourth-order with only {\it two} evaluations of the effective force
(\ref{efff}). 

For three force evaluations, one can use algorithm 4C\cite{chin97}:
\begin{equation}
{\cal T}_{C}^{(4)}(\ep)\equiv
  {\rm e}^{ {1\over 6}\ep T}
  {\rm e}^{ {3\over 8}\ep V+\alpha{1\over{192}}\ep^3 U}
  {\rm e}^{ {1\over 3}\ep T}
  {\rm e}^{ {1\over 4}\ep V+(1-2\alpha){1\over{192}}\ep^3U}
  {\rm e}^{ {1\over 3}\ep T}
  {\rm e}^{ {3\over 8}\ep V+\alpha{1\over{192}}\ep^3U}
  {\rm e}^{ {1\over 6}\ep T},
\label{chinc}
\end{equation}
where $U\equiv [V,[T,V]]$. One is free to distribute the 
commutator term symmetrically via $\alpha$ without affecting 
its fourth-order convergence. The three obvious choices are
$\alpha=0,3/8,1/2$. The first and the last case concentrate 
the gradient term at the center and at the two sides respectively. 
The second case distributes the gradient term in the same
proportion as the original force so that the same effective
force 
\be
F^*(t,E)=[1+{1\over 96}\ep^2 f(t,E)\Bigr]f(t,E)q,
\la{efffc}
\ee
is evaluated at three different times. This is a
direct generalization of algorithm 4B. We shall refer to these
three variants as 4C, 4C$\,^\prime$ and 4C$\,^{\prime\prime}$.
For any specific application, one can fine-tune $\alpha$ to
minimize, or even eliminate, the algorithm's fourth-order 
step-size error. We shall refer to this optimized case as
4C$_{opt}$. Other forward, or just gradient based algorithms, can be found 
in\cite{chin97,chinchen02,chinchen03,ome02,ome03}. 

\section {Killingbeck's backward iteration}

Killingbeck's method\cite{kill85} for solving the eigenvalue-function pair requires
no wave function matching and can be highly automated. It consists of 
two key steps: 1) backward integration to ensure numerical stability, 
and 2) quadratic energy convergence via Newton-Ralphson iterations. 
One begins with an initial guess of the eigenvalue $E^{(0)}$ and chooses a large time 
value $T$ (large $R$ in the original problem) to set
$q(T)=0$ and $p(T)=p_{\infty}$, where $p_{\infty}$ is 
an arbitrary, but small number. One then iterates the algorithm, such as (\ref{alg4b}), 
backward in time to $t=0$. 
(In practice, it may be simpler to run the algorithm
forward in time and change the potential from $V(t)$ to $V(T-t)$.) 
If $E$ is a correct eigenvalue, then it 
must satisfy the eigencondition
\be
q(0,E)=0.
\la{qe}
\ee
Thus the eigenvalue $E$ is a root of the above equation and can be solved by 
Newton-Ralphson iterations:
\be
E^{(n+1)}=E^{(n)}-\frac{q(0,E^{(n)})}{q^{\,\prime}(0,E^{(n)})}.
\la{newton}
\ee 
Killingbeck suggested that the derivative  	
$q^{\,\prime}(0,E)=\partial q(0,E)/\partial E$,
which obeys,
\be
\frac{d^2 q^{\,\prime}(t)}{dt^2}=f(t,E)\,q^{\,\prime}(t)-2q(t),
\la{radpm}
\ee
can be solved simultaneously with $q(t)$, {\it i.e.},  
differentiating any algorithm, such as (\ref{alg4b}),
line by line with respect to $E$.
The resulting algorithm can be iterated at the same time to
determine both $q(0,E)$ and $q^{\,\prime}(0,E)$ simultaneously
so that (\ref{newton}) can be updated directly. By re-running the algorithm
with the updated energy, the procedure can be repeated until convergence.
The convergence is quadratic in the number of iterations. The
converged eigenvalue (and eigenfunction) will deviate from the exact 
value in powers of $\ep^n$ depending on the order of the algorithm used.
 In solving the radial
Schr\"odinger equation, $t=0$ ({\it i.e.}, $r=0$) is the absolute boundary 
and $f(t,E)$ is not defined for $t<0$. Thus in applying
Killingbeck's method, one must not use any algorithm which evaluate the
force at an intermediate time greater than $t+\ep$.

\section {Results for Singular potentials}

One important application of solving the radial Schr\"odinger
equation is in atomic ({\it e.g.}, density functional) calculations, 
where the dominant interaction is the 
Coulomb potential
\be
V(r)=-\frac1{r}.
\la{coulpot}
\ee
As a prototype test case, we show the convergence of
various algorithms in solving for the ground state of
(\ref{coulpot}) in Figure 1. We use $T=26$ ($R=26$); beyond $T=25$, there
is no change in the eigenvalue on the order of $10^{-12}$.
For an initial guess of $E^{(0)}=-0.6$, the Killingbeck
iteration converges to 12 decimal places in 9 iterations or
less. In most cases, once a good guess is found, only a few
iterations are necessary.

It is well known that when the Numerov (N) algorithm\cite{hartree} is
used in Killingbeck's method, the Coubomb ground state only converges 
quadratically\cite{crater}. While the reason for this is 
understood\cite{johnson} and a simple remedy is available\cite{guard}, 
most of the self-starting fourth-order algorithms used here 
suffered no such order reduction. Most can be well-fitted
by the power law $E(\ep)-E_0=c\ep^4$. RKN and RK are 
the three and four force-evaluation Runge-Kutta-Nystrom 
and Runge-Kutta algorithms\cite{battin}, respectively. FR is the Forest-Ruth\cite{forest} 
symplectic algorithm which uses three force-evaluations. This is the first
fourth-order symplectic algorithm found and is well known for its relative
large error. M is McLachlan's improved fourth-order algorithm\cite{mclach} which uses 
four force-evaluations. BM is Blanes and Moan's latest\cite{blanmoan} 
refined fourth-order algorithm which
uses {\it six} force-evaluations.
 
FR, M and BM are examples of conventional symplectic algorithms which have
negative intermendiate time-steps. As shown in Figure 1, algorithm 4B, which use 
only {\it two} evaluations of the effective force,
outperforms all the aforementioned
algorithms except BM regardless the number of force-evaluation. 
OMF18, OMF29, and OMF36 are Omelyan, Mryglod 
and Folk's listed\cite{ome03} fourth-order algorithms 18, 29 and 36.
These are gradient algorithms, similar to 4B and 4C$\,^\prime$, which use
three, four, and five effective force evaluations, respectively. 
As $\alpha$ is varied from 0 to 0.5, the error of the general 4C 
algorithm changes from negative to positive. At $\alpha=0.49$, 
the error curve resembles that of BM. At the optimal value of 
$\alpha=0.41$, the fourth-order error should have been nearly
eliminated with the algorithm showing sixth-order convergence. 
The fact that it does not will be discussed in the next section. 
We fitted all the results in Figure 1
via a power-law of the form $E(\ep)-E_0=c\ep^{n}$ to verify their
order of convergence. All can be well-fitted with $n=4$ except 4B
and 4C$_{opt}$ at $\alpha=0.41$. For 4B, $n\approx 3.5$
and for 4C$_{opt}$, $n\approx 4.5$. Why algorithm 4B should suffer
such an order reduction is not understood. It is possible that 
for 4B, its power-law behavior only sets in at smaller $\ep$. 
The case of 4C$_{opt}$ will be discuss in the next section.
 
Since algorithm OMF29 uses four effective force evaluations, one
can run algorithm 4B twice at the half the time step size. Thus
one should compare OMF29 with 4B($\ep/2$), or 
OMF29(2\,$\ep$) with 4B($\ep$). Thus relative to the computational
effort of 4B($\ep$), one should compare 
4C$\,^\prime$(1.5\,$\ep$), OMF18(1.5\,$\ep$), OMF29(2\,$\ep$),
OMF36(2.5\,$\ep$) and BM(3\,$\ep$). This comparison is shown in Figure 2. 
In this equal effort comparison, algorithm 4B's fourth order
error is as small, if not smaller than all the other 
gradient algorithm's error. This illustrates the case that
efficiency is not necessarily enhanced by increasing the number of 
force evaluations. Also, all gradient algorithms have errors 
smaller than that of BM despite fewer force evaluations. 
We conclude that in solving the Coulomb ground state,
the efficiency of algorithms 4B and 4C$\,^\prime$ 
is unsurpassed by any other algorithms
except by the tunable 4C algorithm.

In the first column of Table I, we list the energy obtained by 
all the algorithms at $\ep=0.01$ weighted by their number of force 
evaluations. Algorithm 4B and 4C$\,^\prime$ indeed turn in the best result
and are outperformed only by 4C$_{opt}$ at 
$\alpha=0.41$. For more accurate results, one can just 
reduce $\ep$. 

As a more stringent test of our algorithms and Killingbeck's
method, we next consider the spiked harmonic oscillator (SHO) 
with potential
\be
V(r)=\frac12 \left(r^2+\frac{\lambda}{r^M}\right).
\la{shopot}
\ee
For extensive references and discussion on SHO, see Refs.\cite{kill01,roy,agu,bue}.
Figure 3 shows the convergence of the ground state energy for the
well studied case\cite{roy,agu,bue} of $M=6$ with $\lambda=0.001$.
For $T=10$ and a reasonable initial guess of $E^{(0)}=1.5$, only
five or less iterations are needed for the energy to converge to 
12 decimal places. For such a singular potential, the convergent 
step size has to be much smaller, but surprisingly, only a magnitude 
smaller. Despite the high degree of singularity, nearly all 
algorithms remained fourth order and none is down graded to lower order. 
At a glance, all 
gradient based algorithms converge better than non-gradient algorithms. 
Even BM is no better than algorithm 4C$\,^\prime$. Since 4C and 
4C$\,^{\prime\prime}$ have errors of opposite sign, one can again 
vary $\alpha$ to minimize the fourth order error. The optimized
algorithm at $\alpha=0.22$ converges better than all other algorithms 
regardless of the number of force evaluations and can be best fitted by
a {\it fifth} order power law. 

To compare the efficiency of gradient algorithms, we again normalize
each algorithm to the computation effort of 4B. In Figure4, we 
plot the convergence curve of 4B($\ep$), 4C$\,^\prime$(1.5\,$\ep$),
OMF29(2\,$\ep$) and OMF36 (2.5\,$\ep$). The solid line
is the fourth order monomial $E(\ep)-E_0=c\ep^4$ which goes through 
4C$\,^\prime$'s result with $c=16.7$. The other three algorithms
can be fitted with the dotted line with $c=20.0$. Thus all gradient
algorithms are essentially similar, with 4C$\,^\prime$ marginally
better. Again, algorithms OME29 and OME36, which use four and five
effective force evaluations with complex numeric coefficients,
are not more efficient than the simpler  
algorithms 4B and 4C$\,^\prime$ with analytical coefficients.

At $\ep=0.001$, the weighted
result of each algorithm is given in the second column of Table I. 
All are in excellent agreement with the value found in the
literature\cite{roy,agu,bue}. At this step size, only gradient algorithms 
are accurate to 10 or more decimal places. For even greater
accuracy, one can simply reduce $\ep$.

The algorithms are equally effective in the case of $M=4$ and 
$\lambda=0.001$. This is shown in Figure 5.
All algorithms showed fourth order convergence, except for 4C$_{opt}$,
which can be better fitted with a fifth order power-law.
Their energy values at $\ep=0.001$ are listed in the third column of
Table I. The optimized 4C algorithm is accurate to 9 decimal
places. Note that once algorithm 4C is optimize for $M=6$,
it can also be used for $M=4$. The change in $\alpha$'s value is
slight.

In the most difficult, ``supersingular" case of $M=5/2$, 
with $\lambda$ remained small at $0.001$,
the power-law behavior seems to require $\ep<10^{-5}$.
This is shown in Figure 6. (If $\lambda$ were not too small, such as
0.1 or 0.01, the power-law behavior would remain observable
in the range of $\ep$ considered.) The energy obtained at
$\ep=0.0002$ is listed in the fourth column of Table I.
The variable 4C algorithm uses $\alpha=0.23$ inherited from
the $M=4$ case. All algorithms are less efficient in dealing with
this ``supersingular" case, but gradient algorithms can still
maintain an 8-digit accuracy. 
   
\section {Higher order algorithms}

In general, if ${\cal T}_A$ is a left-right symmetric approximation of 
the short time evolution operator ${\rm e}^{\,\ep(\hT+\hV)}$,
\be
{\cal T}_A=\prod_{i=1}^N
{\rm e}^{t_i\ep \hT}{\rm e}^{v_i\ep \hV}={\rm e}^{\,\ep H_A}\, ,
\la{arho}
\ee
such that $\sum_i^Nt_i=1$ and $\sum_i^N v_i=1$, then
the approximate Hamiltonian operator is of the form
\bea
\htH &=& T+V\,
+\,\ep^2\,\left(\,\ct\,[T^2\,V]
+\cv\,[V\,T\,V]\,\right)\nn \\
&&+\,\ep^4\,\Bigl(\,\tt\,[T\,T^3\,V]
+ \tv\,[V\,T^3\,V]\nn \\
&&\qquad 
+\,\vt\,[T\,T\,V\,T\,V]\,
+\,\vv\,[V\,T\,V\,T\,V]\,\Bigr)\,+\,\dots\, ,
\la{hop4th}
\eea 
where $\ct$, $\tv$ etc., are coefficients specific to a particular
algorithm and where we have used the condensed commutator notation
$[\hT^2\hV]\equiv[\hT,[\hT,\hV]]$, {\it etc.}. Symmetric factorizations give
rise to time-reversible algorithms and have only even-order error terms.
For a constant $k=\w^2$, the fundamental identity (\ref{ttv}) implies 
that $[TT^3V]=0$ and $[VT^3V]=0$. This crucial simplification 
is no longer true in the time-dependent case when $T$ is replaced 
by $\widetilde T$. From this perspective, one can understand why 
the ability to integrate the 
time-independent harmonic oscillator exactly does not help in solving 
the time-dependent case. Exponentially or sinusoidally fitted 
algorithms are therefore not necessarily more efficient. In the
time-dependent case, the problem is fundamentally different
because some commutators no longer vanish.
Note that the commutators 
\be
[V\widetilde TV\widetilde TV]=4f^2(t)V
\ee 
can be moved back to the LHS. However the saving here is
marginal since this error term can be easily
eliminated by incorporating more gradient terms $[VTV]$
in the factorization process.
 
It has been shown\cite{nosix} that in order to derived a
general sixth-order forward algorithm, one must retain
both $[VTV]$ and $[VT^3V]$ in the factorization process.
Unfortunately, since $[VT^3V]$ cannot be evaluated easily,
there is currently no practical sixth-order forward algorithm.
However, Omelyan, Mryglod and Folk\cite{ome03} have
derived a number sixth-order gradient based, but non-forward,
algorithms of the form
\begin{equation}
{\cal T}_{B}^{(6)}(\ep)\equiv 
\cdots  {\rm e}^{ \ep(v_0V+\ep^2u_0U)}
  {\rm e}^{ \ep t_1 T} 
  {\rm e}^{ \ep(v_1V+\ep^2u_1U)}
  {\rm e}^{ \ep t_2 T}
  {\rm e}^{ \ep(v_2V+\ep^2u_2U)}
  {\rm e}^{ \ep t_3 T}. 
\label{sixb}
\end{equation}
Since the factorization is left-right symmetric, we only
list the operators from the center to the right. These sixth-order 
algorithms all require a mixture of five force or effective force 
evaluations.  
Figure 7 shows the convergence of four of their position-type, 
six-order algorithms
OMF40, OMF41, OMF43 and OMF45 when solving the Coulomb potential. 
None exhibited sixth order convergence. The best is OMF45, which 
converges with power 4.5, same as the optimized, supposedly sixth-order 
algorithm 4C$_{opt}$ with $\alpha=0.41$. Why these sixth-order algorithms
are so down-graded in the Coulomb case is not understood. 

In Figure 8, we compare all algorithms on an equal effort basis as
discussed earlier. In this case, the optimized fourth-order 4C algorithm 
has the smallest error, even when compared with OMF's sixth-order
algorithms.

In Figure 9, we show the convergence of these four sixth-order algorithms in
solving for the ground state energy of the spiked harmonic oscillator with
$M=6$ and $\lambda=0.001$. All OMF algorithms can now be well-fitted 
with sixth order power-laws as indicated by solid lines. In the case of 
OMF40 and OMF41, the ``glitch" in the
convergence curve near $\ep=0.011$ is real. The convergence curve for these 
two algorithms contains a singular term of the form $\approx 1/(\ep-0.011)$,
which blows up near $\ep\approx 0.011$. Why only algorithms
OMF40 and OMF41 exhibit such a singular behavior is also not understood.

In Figure 10, we compare these gradient algorithms in an equal effort basis.
The convergence range of sixth-order algorithms are not 
longer than those of fourth-order algorithms. For $\ep\agt 0.002$, 
the optimized fourth-order algorithm 4C with $\alpha=0.22$ has smaller 
errors than all the sixth-order algorithms. However, for very high 
accuracy, sixth order algorithms are better when $\ep$ is very small.

\section{Conclusions}
In this work, by regarding the radial Schr\"odinger equation as a classical 
time-dependent force problem, we have shown that the entire literature of symplectic 
integrators can be used to find its solution. Among symplectic integrators, factorized
algorithms are favored because Suzuki's rule can be applied
easily to solve the time-dependent force problem. Among
factorized algorithms, gradient or forward algorithms are particularly
suited because they take advantage of the harmonic character
of the Schr\"odinger equation. We demonstrated 
the unique effectiveness of fourth-order gradient symplectic 
algorithms in solving the radial Schr\"odinger equation via
Killingbeck's backward iteration. Even for very singular potentials, these
algorithms are highly effective in computing the eigenvalue-function pair. There
is also no difficulty in obtaining excited states. These gradient 
algorithms can form the core basis for solving non-linear Schr\"odinger equations
such the Hartree-Fock and the Kohn-Sham equation. However, due to the
unique identification of the one-dimensional spatial coordinate as 
time, the current method does not appear to be generalizable to higher 
dimension for solving the general Schr\"odinger equation in 
two- or three-dimension.
 
Among gradient algorithms, algorithm 4C with a tunable parameter $\alpha$
is most efficient in solving a variety of different potentials. 
Despite the fact that there are more complex fourth or sixth-order 
algorithms which use more effective force evaluations, none are really
better than 4C. More force evaluation does not necessarily enhance the
efficiency of an algorithm, specially in solving 
the radial Schr\"odinger equation.

In solving the Coulomb potential, some gradient algorithms are 
down-graded to lower order while others are not. Even more surprising
is the fact that none of the sixth-order algorithm exhibited 
sixth-order convergence in the range of $\ep$ considered.
These findings are not understood and should be studied further.

By regarding the radial Schr\"odinger equation as a classical 
dynamical problem, one can now use the same set of symplectic 
algorithms for solving both classical and quantum mechanical problems.

\begin{acknowledgments}
This work was supported in part, by a National Science Foundation 
grant (to SAC) No. DMS-0310580 and ONR grant No. N00014-04-1-0336
to Marlan Scully.
\end{acknowledgments}
\bigskip
\bigskip
\bigskip
\centerline{REFERENCES}

\begin{table}[t]
\caption[]{ 
Equal computational effort comparison of all fourth order
algorithms.  
}
\begin{center}
\begin{tabular}{|c|c|c|c|c|} 
\hline  
           &Coulomb     &SHO(M=6)        &SHO(M=4)   &SHO(M=5/2) \\  
           &$\ep=0.01$  &$\ep=0.001$     &$\ep=0.001$ & $\ep=0.0002$     \\ \hline
4B($\,\ep$)           &\,-0.49999999968\,   &\,1.63992791294\, &\,1.53438158386\,&1.502005640    \\ \hline
FR(1.5$\,\ep$)            &-0.49999973551     &1.63992789976     &1.53438108257  &1.502005464         \\ \hline
RKN(1.5$\,\ep$)           &-0.50000000543     &1.63992791316     &1.53438159696  &1.502005154   \\ \hline
4C$\,^\prime$(1.5$\,\ep$) &-0.50000000020     &1.63992791294     &1.53438158417  &1.502005637   \\ \hline
4C$_{opt}$(1.5$\,\ep$)    &-0.50000000005     &1.63992791296     &1.53438158529  &1.502005613  \\ \hline
OMF18(1.5$\,\ep$)         &-0.49999999943     &1.63992791294     &1.53438158351  &1.502005644      \\ \hline
RK(2$\,\ep$)              &-0.49999996158     &1.63992790762     &1.53438117625  &1.502004936   \\ \hline
M(2$\,\ep$)               &-0.50000000924     &1.63992791131     &1.53438145329  &1.502005456    \\ \hline
OMF29(2$\,\ep$)           &-0.49999999952     &1.63992791294     &1.53438158387  &1.502005640    \\ \hline
OMF36(2.5$\,\ep$)         &-0.49999999967     &1.63992791294     &1.53438158389  &1.502005579    \\ \hline
BM(3$\,\ep$)              &-0.49999999791     &1.63992791269     &1.53438156400  &1.502005496    \\ \hline
\hline
``Exact"\cite{roy,agu,bue}    &-0.50000000000     &1.63992791296     &1.53438158545  &1.502005626      \\ \hline
\end{tabular}
\end{center}
\label{tabal}
\end{table}
\newpage
\begin{figure}
	\vspace{0.5truein}
	\centerline{\includegraphics[width=0.8\linewidth]{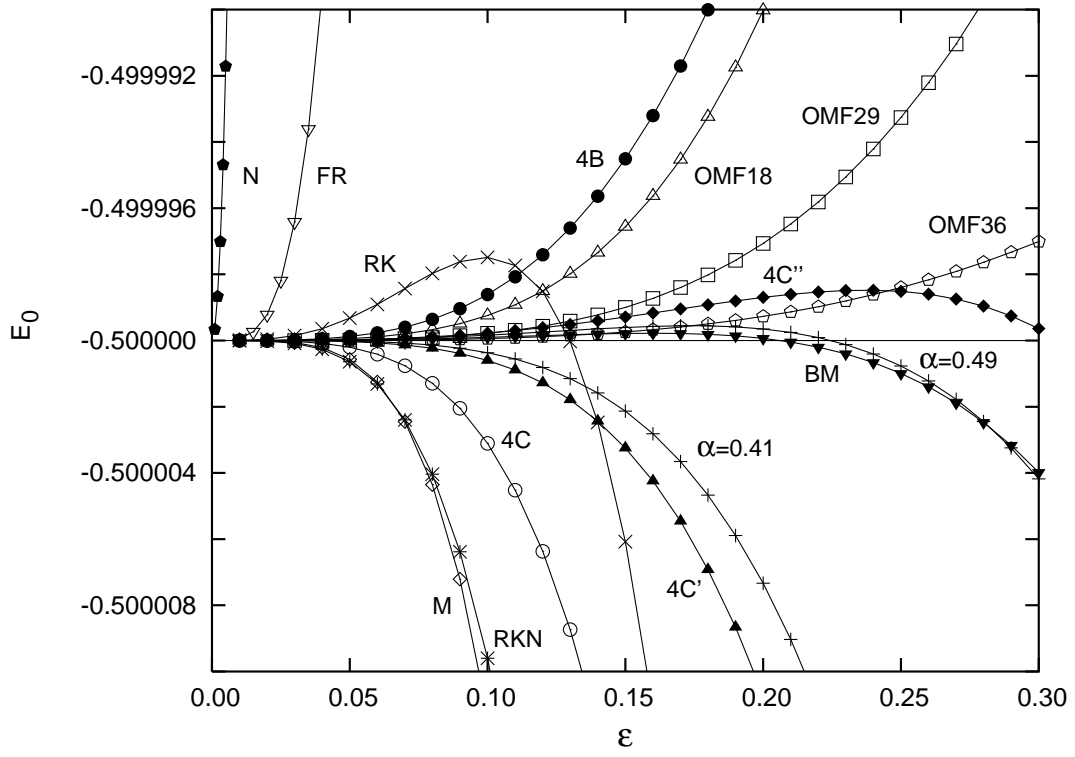}}
	\vspace{0.5truein}
\caption{The convergence of various fourth order algorithms in solving
for the ground state energy of the Coulomb potential. Solid lines only
connect data points to guide the eye. See text for identification
of each algorithm.  
\label{fig1}}
\end{figure}
\begin{figure}
	\vspace{0.5truein}
	\centerline{\includegraphics[width=0.8\linewidth]{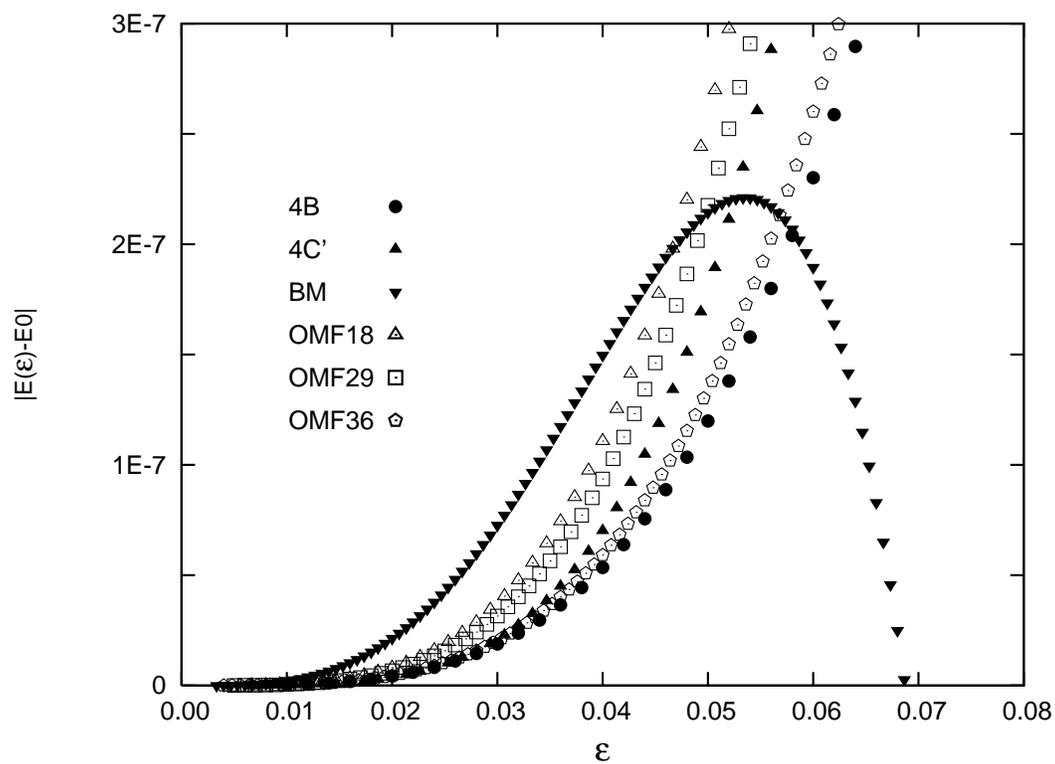}}
	\vspace{0.5truein}
\caption{Equal computational effort comparison of selected algorithms in solving
for the Coulomb ground state energy. See text for discussion.  
  \label{fig2}}
\end{figure}
\begin{figure}
	\vspace{0.5truein}
	\centerline{\includegraphics[width=0.8\linewidth]{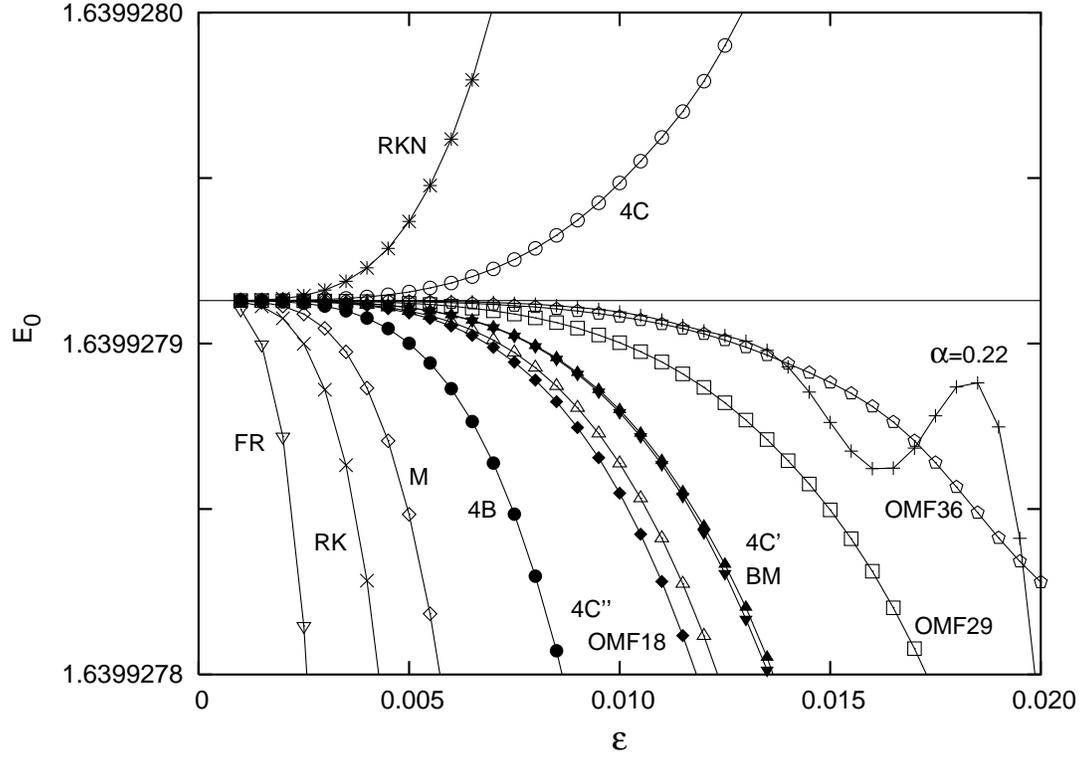}}
	\vspace{0.5truein}
\caption{The convergence of various fourth-order algorithms in solving for the ground state 
energy of the spiked harmonic oscillation (\ref{shopot}) with $M=6$ and $\lambda=0.001$. Same
plotting symbols are used to designate the same algorithm compared in Figure \ref{fig1}. 
  \label{fig3}}
\end{figure}
\begin{figure}
	\vspace{0.5truein}
	\centerline{\includegraphics[width=0.8\linewidth]{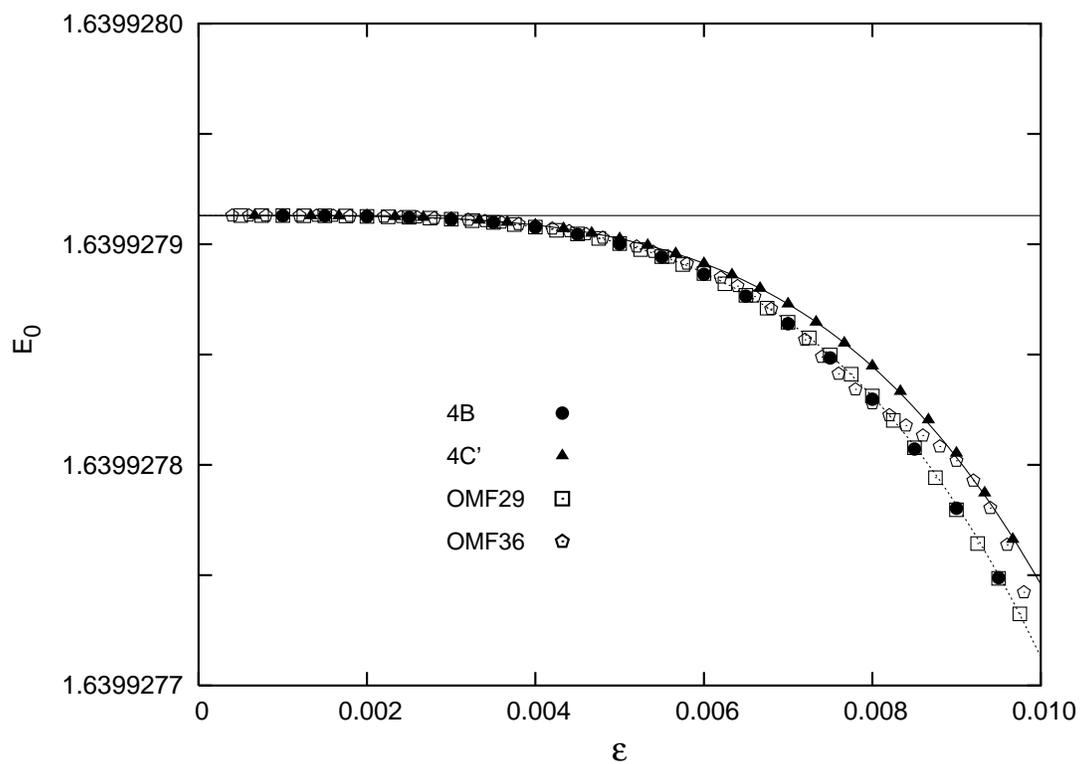}}
	\vspace{0.5truein}
\caption{Equal effort comparison of various fourth-order gradient symplectic algorithms in
solving for the ground state energy of the spiked harmonic oscillator of Figure \ref{fig3}.  
  \label{fig4}}
\end{figure}
\begin{figure}
	\vspace{0.5truein}
	\centerline{\includegraphics[width=0.8\linewidth]{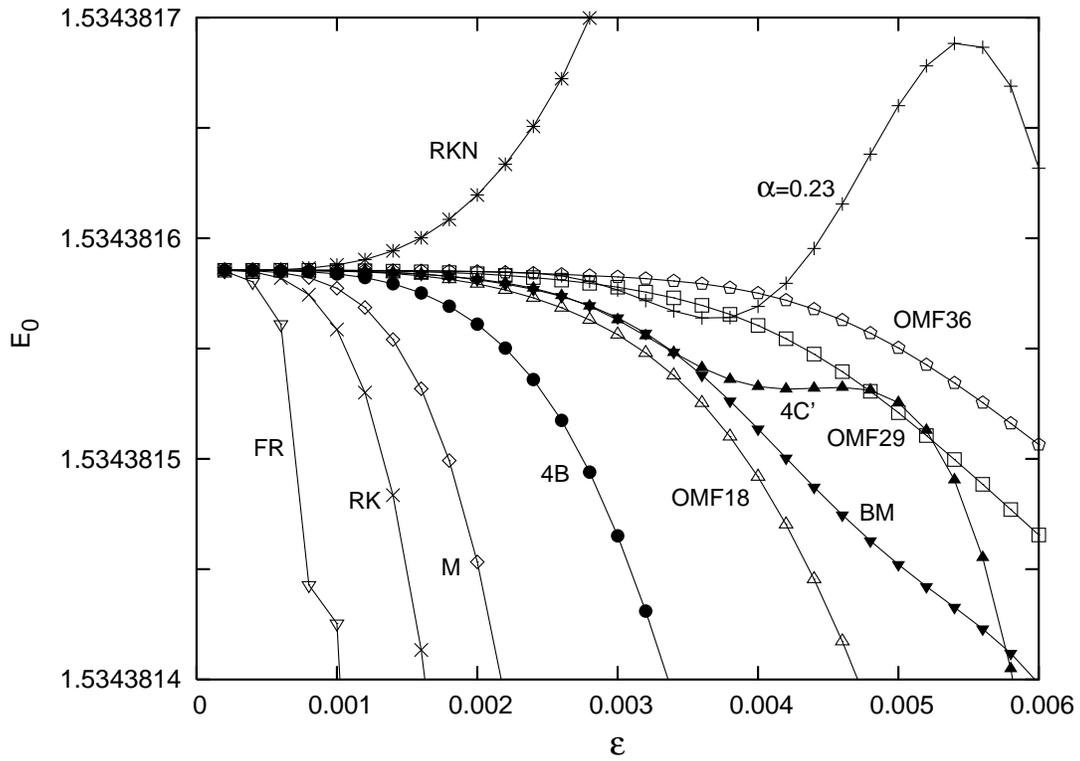}}
	\vspace{0.5truein}
\caption{The convergence of various fourth-order algorithms in solving for the
ground state energy of the spiked harmonic oscillation (\ref{shopot}) with $M=4$ 
and $\lambda=0.001$. 
  \label{fig5}}
\end{figure}
\begin{figure}
	\vspace{0.5truein}
	\centerline{\includegraphics[width=0.8\linewidth]{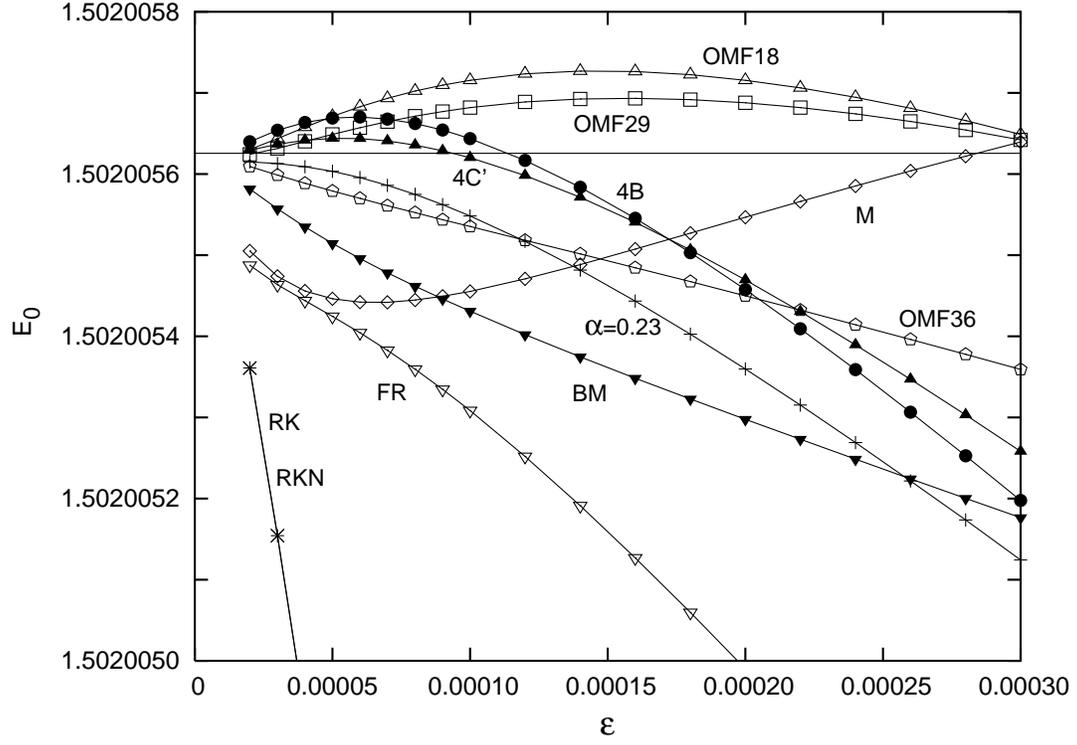}}
	\vspace{0.5truein}
\caption{The convergence of various fourth-order algorithms in solving for the
ground state energy of the spiked harmonic oscillation (\ref{shopot}) with $M=5/2$ 
and $\lambda=0.001$.
In this ``supersingular" case, the power-law behavior is not observed within the
range of $\ep$ considered.
  \label{fig6}}
\end{figure}
\begin{figure}
	\vspace{0.5truein}
	\centerline{\includegraphics[width=0.8\linewidth]{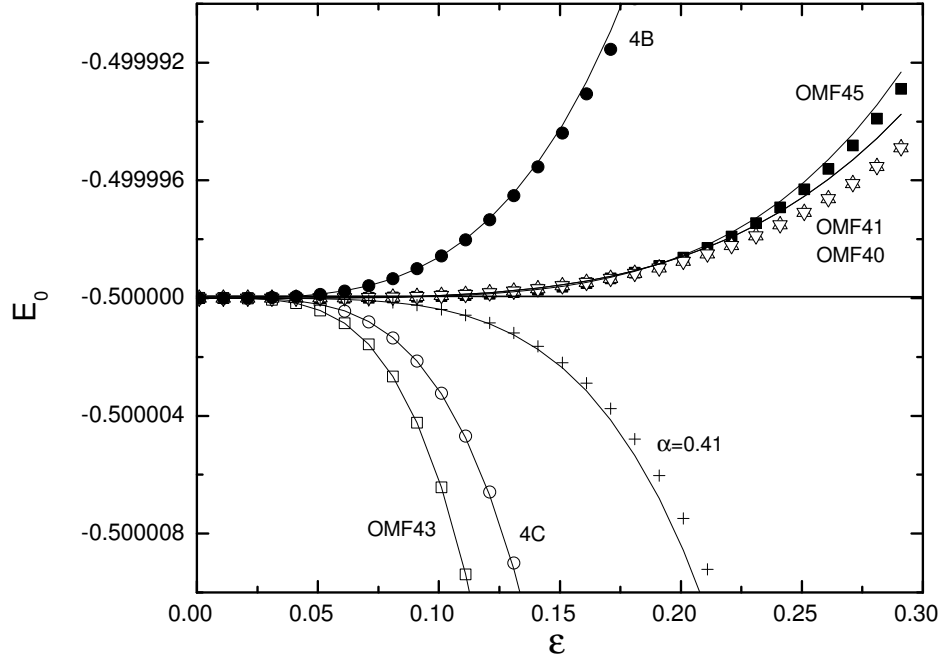}}
	\vspace{0.5truein}
\caption{The convergence of various fourth and sixth order gradient 
algorithms in solving for the
ground state energy of the Coulomb potential. The solid lines here are 
fitted power-laws of power 3.5 (4B), 4 (4C, OMF40, OMF41, OMF43),
and 4.5 (4C with $\alpha=0.41$, OMF45). None is showing sixth order 
convergence. 
  \label{fig7}}
\end{figure}
\begin{figure}
	\vspace{0.5truein}
	\centerline{\includegraphics[width=0.8\linewidth]{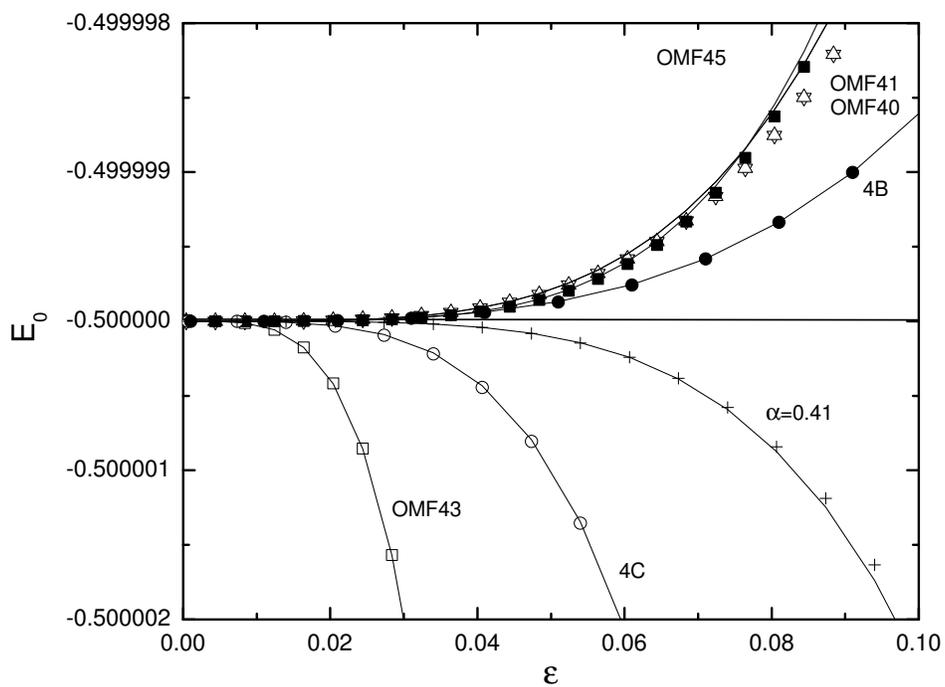}}
	\vspace{0.5truein}
\caption{Equal effort comparison in solving for the
ground state energy of the Coulomb potential. The OMF algorithms are nominally
sixth-order algorithms. However, their convergence is no better than
that of algorithm 4C with $\alpha=0.41$.
  \label{fig8}}
\end{figure}
\begin{figure}
	\vspace{0.5truein}
	\centerline{\includegraphics[width=0.8\linewidth]{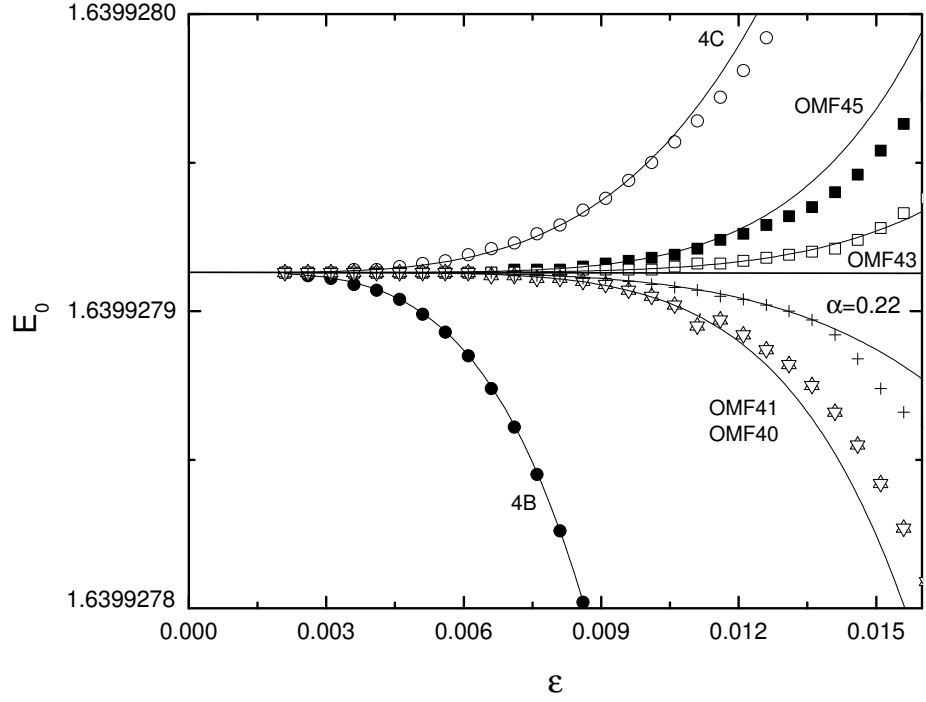}}
	\vspace{0.5truein}
\caption{The convergence of various fourth and sixth-order algorithms in solving 
for the ground state energy of the
spiked harmonic oscillator (\ref{shopot}) with $M=6$ and $\lambda=0.001$. 
The solid lines are fitted 
power-laws of power 4 (4B, 4C), 5 (4C with $\alpha=0.22$),
and 6 (all OMF algorithms). 
  \label{fig9}}
\end{figure}
\begin{figure}
	\vspace{0.5truein}
	\centerline{\includegraphics[width=0.8\linewidth]{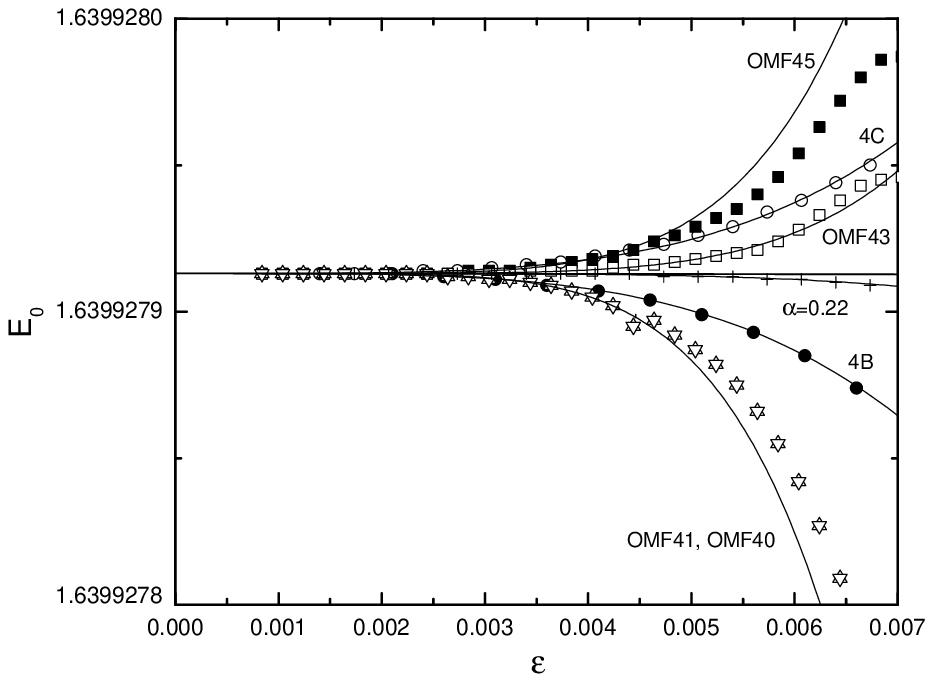}}
	\vspace{0.5truein}
\caption{Equal effort comparison of various fourth and sixth-order gradient 
symplectic algorithms in solving for the ground state energy of the spiked 
harmonic oscillator of Figure \ref{fig9}. The optimized algorithm 4C 
with $\alpha=0.22$ has
the smallest error for $\ep\agt 0.002$. 
  \label{fig10}}
\end{figure}

\begin{thebibliography}{}
\bibitem{hartree}D. R. Hartree, {\it The Calculation of Atomic Structures},
                 Pergamon, London, New York, 1957.
\bibitem{rchen}R. Chen, Z. Xu, and Lan Sun,
                Phys. Rev. E {\bf 47}, 3799 (1993).
\bibitem{raptis} A. Raptis and A. C. Allison, Comp. Phys. Comm. {\bf 14}, 1 (1978).
\bibitem{simos}	T. E. Simos, IMA J. Numer. Anal. {\bf 21}, 919 (2001).
\bibitem{vyver} H. Van de Vyver, Comp. Phys. Comm. {\bf 166}, 109 (2005).
\bibitem{battin}R. H. Battin {\it An Introduction to the
          Mathematics and Methods of Astrodynamics, Reviesed Edition},
		  AIAA, 1999.
\bibitem{dragt} A. J. Dragt and J. M. Finn,
       J. Math. Phys. {\bf 17}, 2215 (1976)
\bibitem{yoshi} H. Yoshida, Celest. Mech. {\bf 56}, 27 (1993).
\bibitem{mcl02} R. I. McLachlan and G. R. W. Quispel, Acta Numerica,
               {\bf 11}, 241 (2002).
\bibitem{hairer}E. Hairer, C. Lubich, and G. Wanner, 
               {\it Geometric Numerical Integration}, 
                Springer-Verlag, Berlin-New York, 2002.
\bibitem{gla91}B. Gladman, M. Duncan, and J. Candy, 
            Celest. Mech. Dyn. Astron. {\bf 52}, 221 (1991).
\bibitem{chin97} S. A. Chin, Phys. Lett. A {\bf 226}, 344 (1997).
\bibitem{chinkid}S. A. Chin and D. W. Kidwell,
                Phys. Rev. E {\bf 62}, 8746 (2000).
\bibitem{chinchen03}S. A. Chin, and C. R. Chen,
         Cele. Mech. and Dyn. Astron. {\bf 91}, 301 (2005)
\bibitem{chinsante}S. Scuro and S. A. Chin, Phys. Rev. E {\bf 71}, 056703 (2005).
\bibitem{liu} X.S. Liu, X.Y. Liu, Z.Y. Zhou, P.Z. Ding and S.F. Pan,
              Int. J. Quant. Chem {\bf 79}, 343 (2000).
\bibitem{kalog} Z. Kalogiratou, Th. Monovasilis and T.E. Simos,
              J. Comput. Appl. Math. {\bf 158}, 83 (2003).
\bibitem{wensch} J. Wench, M. D\"ane, W. Hergert and A. Ernst, 
                 Comp. Phys. Comm. {\bf 160}, 129 (2004).
\bibitem{baye} D. Baye, G. Goldstein and P. Capel, 
         Phys. Lett. A {\bf 317}, 337 (2003)
\bibitem{gold} G. Goldstein and D. Baye, 
                Phys. Rev. E {\bf 70}, 056703 (2004).
\bibitem{suz93}M. Suzuki, Proc. Japan Acad. Ser. B, {\bf 69}, 161 (1993).
\bibitem{chinchen02}S. A. Chin and C. R. Chen,
                J. Chem. Phys. {\bf 117}, 1409 (2002).
\bibitem{ome02}I. P. Omelyan, I. M. Mryglod and R. Folk,
               Phys. Rev. E {\bf 66}, 026701 (2002).
\bibitem{ome03}I. P. Omelyan, I. M. Mryglod and R. Folk,
               Comput. Phys. Commun. {\bf 151}, 272 (2003)
\bibitem{chin2001}S. A. Chin, and C. R. Chen, 
		  J. Chem. Phys. {\bf 114}, 7338 (2001).
\bibitem{chinkro} S. A. Chin and E. Krotscheck,
		  Phys. Rev. E {\bf 72}, 036705 (2005).
\bibitem{kill85} J. Killingbeck, J. Phys. A {\bf 18}, 245 (1985).
\bibitem{suzfour}M. Suzuki, {\it Computer Simulation Studies in 
            Condensed Matter Physics VIII},
           eds, D. Landau, K. Mon and H. Shuttler (Springler, Berlin, 1996).
\bibitem{crater}H.W. Crater and G.W. Redden, 
		  J. Compt. Phys. {\bf 19}, 236 (1975).
\bibitem{johnson} B. R. Johnson, 
		  J. Chem. Phys. {\bf 69}, 4676 (1978).
\bibitem{guard}E. Buendia and R. Guardiola, 
		  J. Compt. Phys. {\bf 60}, 561 (1985).
\bibitem{forest} E. Forest and R. D. Ruth, 
           Physica D {\bf 43}, 105 (1990).
\bibitem{mclach} R. I. McLachlan, SIAM J. Sci. Comput. {\bf 16}, 151 (1995).
\bibitem{blanmoan} S. Blanes and P.C. Moan, as quoted in Ref.9, P.407,
                  algorithm 1(a).
\bibitem{kill01} J. P. Killingbeck, G. Jolicard and A. Grosjean, 
		  J. Phys. A {\bf 34}, L367 (2001).
\bibitem{roy}A. K. Roy, 
		  Phys. Lett. A {\bf 321}, 231 (2004).
\bibitem{agu} V. C. Aguilera-Navarro, G. A. Est\'evez and R Guardiola,
              J. Math. Phys. {\bf 31}, 99 (1990).
\bibitem{bue} E. Buend\'ia, F. J. G\'alvez and A Puertas, J. Phys. A {\bf 28}, 6731 (1995).
\bibitem{nosix} S. A. Chin, Phys. Rev. E {\bf 71}, 016703 (2005).

\end{thebibliography}
\end{document}